# Tunable Electronic Structure and Magnetic Coupling in Strained Two-Dimensional Semiconductor MnPSe$_3$

Qi Pei[1], Xiaocha Wang[2], Jijun Zou[3], Wenbo Mi[1,*]

[1]*Tianjin Key Laboratory of Low Dimensional Materials Physics and Preparation Technology, School of Science, Tianjin University, Tianjin 300354, China*

[2]*School of Electrical and Electronic Engineering, Tianjin University of Technology, Tianjin 300384, China*

[3]*Key Laboratory for Green Chemical Technology of the Ministry of Education, School of Chemical Engineering and Technology, Tianjin University, Tianjin 300354, China*

[*]Author to whom all correspondence should be addressed.

E-mail: miwenbo@tju.edu.cn

# ABSTRACT

The electronic structures and magnetic properties of strained monolayer MnPSe$_3$ are investigated systematically by first-principles calculations. It is found that the magnetic ground state (GS) of monolayer MnPSe$_3$ can be significantly affected by biaxial strain engineering, while the semiconducting characteristics are well preserved. Owing to the sensitivity of the magnetic coupling towards the structural deformation, a biaxial tensile strain about 13% can lead to an antiferromagnetic-ferromagnetic (AFM-FM) transition. The underlying physical mechanism of strain-dependent magnetic stability is mainly attributed to the competition effect of direct AFM interaction and indirect FM superexchange interaction between the nearest-neighbor (NN) two Mn atoms. In addition, we find that FM MnPSe$_3$ is an intrinsic half semiconductor with a large spin exchange splitting in conduction bands, which is crucial for the spin-polarized carrier injection and detection. The sensitive interdependence among external stimuli, electronic structure and magnetic coupling suggests that monolayer MnPSe$_3$ can be a promising candidate in spintronics.

# 1 Introduction

Recently, the emergence of two-dimensional (2D) materials has aroused widespread concerns [1-4]. Especially, 2D magnetic semiconductors who possess both semiconducting and magnetic characteristics have been considered as promising candidates for low-dimensional spintronic devices. In general, magnetic semiconductors can be classified into two categories: one is intrinsic magnetic semiconductor and the other is diluted magnetic semiconductor. Compared with intrinsic magnetic semiconductor, the diluted magnetic semiconductor can be easily obtained by doping the nonmagnetic semiconductor with transition metals [5-7]. However, there still exist some problems in synthesis, such as low solubility and rich boundary defects. Therefore, searching for the experimentally feasible methods for the fabrication of magnetic semiconductors is full of importance. At present, the strategy of exfoliating single layers from bulk layered structures has been comprehensively investigated. Following this approach, recent theoretical studies have proposed some potential magnetic 2D van der Waals (vdW) materials, including transition metal carbides and nitride *MX*enes [8], chromium-based ternary tritellurides Cr*X*Te$_3$ (*X*=Si, Ge) [9-11], vanadium-based dichalcogenides V*X*$_2$ (*X*=S, Se) [12], trihalides Cr*X*$_3$ (*X*=F, Cl, Br, I) [13] and transition metal phosphorus trichalcogenides *M*P*X*$_3$ (*M*=Fe, Mn, Ni; *X*=S, Se) [14-16]. As one of the representative in *MPX*$_3$ family, monolayer MnPSe$_3$ exhibits particular advantages in solar-energy-related applications and novel spin-valley coupling physics [17, 18], which endows this 2D crystal with great potentials in spintronics.

The control of the spin ordering is another key issue for the application of spintronic devices. The spin ordering of magnetic materials should be deliberately modulated by external stimuli, such as electric field, atomic vacancy, adsorption, doping, or elastic strain engineering [19-22]. Among these external factors, mechanical strain is commonly regarded as an effective route to induce the change of crystal structure, the modification of orbital overlap and the emergence of charge distribution. Meantime, the excellent mechanical flexibility of 2D magnets provides further proof of the strain-engineering's feasibility [23, 24]. Many theoretical reports have shown the tunable electronic structures and magnetic characteristics in stained monolayers [12, 25-28]. Experimentally, a remarkable advance in applying tunable biaxial strain to 2D materials has been proposed by Ding *et al.* [29]. Using a thin layer of polymethyl methacrylate (PMMA) as glue, single layer graphene can be transferred onto a $SiO_2$ covered piezoelectric substrate. Then a bias voltage applied to the piezoelectric substrate will result in an out-of-plane electric field and subsequently lead to an in-plane strain. This technique not only makes it possible to study strain-related behaviors of low-dimensional materials but also inspires new ideas and methods in strain application.

In this paper, the electronic and magnetic properties of monolayer $MnPSe_3$ are investigated by density functional theory systemically, as well as the intercoupling between the strain and magnetism. It is found that the pristine monolayer $MnPSe_3$ is an intrinsic semiconductor with antiferromagnetic (AFM) ordering. The stability of AFM coupling can be significantly weakened by applying the elastic biaxial strains. An antiferromagnetic-ferromagnetic (AFM-FM) transition occurs under a large tensile strain beyond 13%. The physical mechanism of such a phenomenon is further studied, and a strain-related competition between direct exchange interaction and indirect superexchange interaction is put forward. In addition, monolayer MnPSe$_3$ with FM coupling exhibits robust semiconducting characteristics, with fully spin-polarized valance and conduction band edges. These novel properties render monolayer MnPSe$_3$ a promising platform to explore magnetic phenomena and offer a promising avenue for fabricating controllable and tunable spintronic devices.

## 2 Computational details

Our calculations are performed in the framework of density functional theory (DFT), using a plane-wave basis [30] set as implemented in the VASP code [31]. The generalized gradient approximation (GGA) parameterized by Perdew-Burke-Ernzerhof (PBE) [32] is utilized to describe exchange-correlation functional. For the reason that GGA cannot properly describe the strongly correlated systems with partially filled $d$ subshells, we use Hubbard $U$ terms (5 eV for Mn) to describe the on-site electron-electron Coulomb repulsion as suggested in the literature [33]. The vdW-D2 correction [34] is used together with GGA+$U$ to add the longer-ranged correlation in evaluating vdW interaction between the monolayers. Kohn-Sham single-particle wavefunctions are expanded in the plane wave basis set with a kinetic energy truncation at 500 eV. The Monkhorst-Pack scheme is adopted for $k$-point sampling employing a 7×7×1 $\Gamma$-centered grid. The energy and force convergence criteria on each atom are less than $10^{-6}$ eV and 0.01 eV/Å, respectively. A 2×2×1 supercell model including eight Mn atoms is adopted in calculation to indentify the preferred magnetic ground state (GS), and a 20 Å vacuum slab is inserted perpendicularly on the top of MnPSe$_3$ surface to minimize the interaction between periodic images.

## 3 Results and discussion

Bulk MnPSe$_3$ has been proposed to be an AFM semiconductor with the optical energy gap of 2.27 eV [35] and Neél temperature of 74 K [36]. Since the relatively weak vdW force holds together the layered compound, monolayer MnPSe$_3$ is easily exploited from its corresponding bulk phase. As shown in Figs. 1(a) and (b), each MnPSe$_3$ unit cell is composed of two Mn$^{2+}$ ions and one [P$_2$Se$_6$]$^{4-}$ cluster. A Mn atom locates at the center of a distorted octahedron with six Se atoms arranged in a trigonal antiprism structure. The dumbbell-like P dimer in [P$_2$Se$_6$]$^{4-}$ bipyramid constrains six neighboring Se atoms to form two Se trimers with a relative in-plane twist of 60°, perpendicular to the honeycomb plane. The formal valance of Mn, P and Se atoms are +2, +4 and -2, respectively. According to Hund's rule, Mn$^{2+}$ with a 3$d^5$ electronic configuration will exhibit a high spin state.

Lattice structure of MnPSe$_3$ monolayer is optimized to observe the resulting lattice parameters. After fully relaxation, the lattice constant of the FM structure (6.409 Å) is quite similar to the AFM (6.403 Å), suggesting the irrelevance between the lattice parameter and magnetic structure. The agreement of optimized lattice constant with the experimental value (6.387 Å) [36] also reveals that GGA+$U$ can correctly describe the structural and electronic properties of monolayer MnPSe$_3$, verifying the rationality of our parameters in calculation. The spin density for FM and AFM configurations are calculated by the equation of $\rho = \rho_\uparrow - \rho_\downarrow$, where the $\rho_\uparrow$ and $\rho_\downarrow$ represent spin-up and spin-down charge densities, respectively. As illustrated in Fig. 1(c), the major magnetism in monolayer MnPSe$_3$ with FM ordering is contributed by Mn atoms. The local magnetic moment of Mn atom is textured to be about 4.613 $\mu_B$, and each P and Se atom carries about 0.016 and 0.023 $\mu_B$ magnetic moments. Similarly, the spin density for AFM configuration in Fig. 1(d) also concentrates on two Mn atoms with opposite spin magnetic moment (4.598 $\mu_B$), while the magnetic moment of each P and Se can be neglected. The total magnetic moment of AFM form is exactly 0 $\mu_B$ for the intrinsic AFM coupling of Mn$^{2+}$ ions.

To confirm the most preferable magnetic coupling between the Mn atoms, with the exception of FM and regular AFM (Néel) orders, we consider two additional AFM configurations, which are zigzag-AFM and stripy-AFM shown in Fig. 1(e). After differentiate them energetically, the relative energy differences Δ$E$ (absolute value) with respect to the Néel-AFM configuration are 0, 37.77, 13.12 and 17.92 meV for Néel-AFM, FM, zigzag-AFM and stripy-AFM, respectively. Our calculations clearly show that the Néel-AFM order has the minimum energy, indicating the GS of Néel order. Unless indicated otherwise, AFM order will refer to the AFM-Néel order in the following. Besides, the non-magnetic state can be neglected for the great energy disparity between the NM state and magnetic states.

To evaluate the effect of strain engineering on the electronic and magnetic properties, a series of *xy*-plane biaxial strains are applied to monolayer MnPSe$_3$. As shown in Fig. 2(c), we stretch or shrunk the lattice constants uniformly to simulate the tensile and compressive strains. The strain can be estimated by considering the lattice constants quantitatively as $\varepsilon = (a - a_0)/a_0$, where $a_0$ and *a* are the lattice constants of the original and strained models, respectively. A distinction between tensile and compressive strains is made via defining $\varepsilon>0$ or $\varepsilon<0$. The variation of energy difference between the AFM and FM coupling in biaxial strained systems is depicted in Fig. 2(a). With the tensile strain changes from 0% to 5%, the energy difference $\Delta E_{\text{AFM-FM}}$ increases from -37.77 to -20.30 meV per formula unit, suggesting that the AFM state is significantly weakened and tends towards instability. When the tensile strain continues to increase, a clear magnetic phase transition from AFM to FM is achieved in the presence of a relatively large strain about 13%. Moreover, the FM stability can be further enhanced under a 15% tensile strain. Since large strain modulations (beyond 10%) have already been reported in many ultrathin 2D systems, such as graphene [37] and MoS$_2$ [25] by using either the three-point bending configuration or piezoelectric substrates, we are quite right in believing that 13% tensile strain can also be realized in monolayer MnPSe$_3$. Hence, strain engineering can be considered as a potential modulating artifice to realize the transition of magnetic phase in monolayer MnPSe$_3$. Meanwhile, the distance between the nearest-neighbor (NN) two Mn atoms increases linearly from 3.515 to 4.267 Å as the biaxial strain varying from -5% to 15%. Actually, we calculate the total energies for all possible magnetic configurations under each biaxial strain, and the variation of energy difference ($\Delta E_{\text{AFM-zigzag}}$ and $\Delta E_{\text{AFM-stripy}}$) is shown in Fig. S1 in the Supplementary Information. Since there is no other magnetic phase transition occurring within the biaxial strain range we consider, we only focus on the AFM to FM transition.

Fig. 2(b) shows the variation of band gap in biaxial strained systems. Within the strain region from -5% to 3%, the band gap increases gradually. However, when the tensile strain increases continuously, band gap values begin to decrease. A sharp band gap shrinking from 1.09 to 0.67 eV appears at the tensile strain of 13%, corresponding to the magnetic phase transition from AFM to FM. In addition, the robustness of semiconducting property against wide range of biaxial strain intensity also should be noted.

Band structure (BS) and density of states (DOS) within the energy window ranging from -3 to 3 eV are calculated to reveal the electronic properties in strained monolayer $MnPSe_3$. Figs. 3(a)-(f) denote to the systems under -5%, 0%, 5%, 10%, 13% and 15% strains, respectively. For the pristine monolayer (see Fig. 3(b)), $MnPSe_3$ is a semiconductor with a direct band gap of 1.83 eV, in accordance with previous theoretical results [14, 17]. The valence bands (VB) near Fermi level mainly consist of Se and Mn atom states, whereas the conduction bands (CB) are contributed by Se, Mn, and P atoms. A comparison of Figs. 3(c) and (e) shows that, as the lattice constants increase, the contribution of the Se $p$ orbitals to the states in the range of VBM to -0.5 eV increases. The peak of the Se PDOS in this energy range shifts to higher energies. Moreover, by applying a 5% tensile strain, BS calculations reveal an indirect semiconductor with the conduction band minimum and valance band maximum at $\Gamma$ and $K$ points, respectively. The band transition from direct to indirect also makes the gap decrease to 1.64 eV, which is well consistent with the variation tendency given in Fig. 2(b). With the strain increased to 13%, FM state is more stable than AFM state. The occupied states near Fermi level have major contribution from Se atoms, while the unoccupied states in the vicinity of Fermi level are deriving from Se, Mn, and P atoms. Examination of the PDOS of Mn and Se atoms shows that Mn $d$ states are relatively delocalized and hybridized with Se $p$ states, indicating that indirect $p$-$d$ exchange interaction plays an important role in the ferromagnetic coupling. In addition, FM $MnPSe_3$ is an indirect semiconductor with the band gap of 0.66 eV. The bands in the spin-up and spin-down channels are not overlapped, suggesting the intrinsic ferromagnetism. Particularly, the VB and CB edges of FM MnPSe$_3$ are fully spin-polarized in the same spin direction, indicating that the FM MnPSe$_3$ is a half semiconductor (HSC). A large spin exchange splitting of 0.40 eV (labled as Δ in Fig. 3(e)) in the CB is observed, which is not only essential for the spin-polarized carrier injection and detection [10], but also crucial for gaining half-metallic property by shifting the relative position of Fermi level. The exchange splitting can be further increased to 0.44 eV at a tensile strain of 15%.

Because our calculations show that the magnetic moments are concentrated on the metal atoms sites (see Fig. 1(c)-(d)), we characterize the magnetic properties of MnPSe$_3$ with an effective Heisenberg model Hamiltonian on a honeycomb lattice,

$$H = \sum_{<ij>} J_1 \mathbf{S}_i \cdot \mathbf{S}_j + \sum_{<ij>} J_2 \mathbf{S}_i \cdot \mathbf{S}_j + \sum_{<ij>} J_3 \mathbf{S}_i \cdot \mathbf{S}_j \qquad (1)$$

where $J_{1,2,3}$ are the exchange interactions between NN, second NN, and third NN spins. $\mathbf{S}_i$ is the total spin magnetic moment of the atomic site $i$. To further extract the magnetic exchange interactions, the lattice is fixed to that of the most energetically favorable spin confirmation and the energies for different spin configurations are computed. Using equation (1), the magnetic energy can be explicitly expressed as

$$E_{\text{FM/Néel}} = E_0 + (\pm 3J_1 + 6J_2 \pm 3J_3)|\mathbf{S}|^2 \qquad (2)$$

$$E_{\text{zigzag/stripy}} = E_0 + (\pm J_1 - 2J_2 \mp 3J_3)|\mathbf{S}|^2 \qquad (3)$$

We then obtain the lattice constant, GS along with the exchange coupling constants for pristine and biaxial strained MnPSe$_3$. As listed in Table S1, the NN interaction $J_1$, the second NN interaction $J_2$ and the third NN interaction $J_3$ are all AFM. The value of $J_2$ (0.02 meV) is one magnitude less than $J_1$ (0.195 meV) and much smaller than $J_3$ (0.103 meV). These findings are consistent with previous studies on MnPS$_3$ and MnPSe$_3$ [16, 38, 39]. More importantly, $J_2$ and $J_3$ interactions in transition-metal trichalcogenide monolayers are always AFM [40].

To understand the microscopic origin of the exchange interactions in monolayer MnPSe$_3$, the possible electron hopping paths for $J_1$, $J_2$ and $J_3$ interactions are plotted. As shown in Fig. 4(a), electrons hopping for $J_1$ interaction mainly go through two paths. One is the short-range direct interaction between two neighboring Mn ions (Mn-Mn), which is robustly AFM for the half-filled high-spin $d^5$ state of Mn, and the other is the more long-range Mn-Se-Mn superexchange interaction with an angle of 84.1º. According to the well-known Goodenough-Kanamori-Anderson (GKA) rules [41, 42], systems with cation-anion-cation bond angles of 90° prefer weak FM ordering and 180° superexchange interactions are AFM. Therefore, the Mn-Se-Mn angle (84.1º) close to 90° in superexchange is FM. Due to the closed $d$ shell on Mn ions and large electron excitation energy from Se $p$ orbital to Mn $d$ orbital, the direct AFM exchange interaction dominates over FM superexchange interaction. As a consequence, $J_1$ is expected to be AFM.

In Figs. 4(b) and (c), there are several hopping paths for $J_2$ and $J_3$, where the cations are separated by two anions. Based on the geometry, the most possible paths should involve two se anions on the same plane with the shortest distance. For this reason, $J_2$ and $J_3$ can be regarded as super-superexchange interactions. Since the rules for cation-anion-anion-cation path are similar to those for cation-anion-cation interactions, 180° and 90° are also two criterion angles for AFM and FM interactions, respectively. However, for those interactions with intermediate angels between 90° and 180°, a crossover angle of 127±0.6° is related to the AFM-FM transition [43]. Zhang *et al*. [44]

also proposed that a large cation-anion-anion angle beyond 130° leads to AFM interaction, while an angel about 90° contributes an FM interaction. According to the above mentioned, we find that the most possible electron hopping path for $J_3$ contains two cation-anion-anion angels, which are both 132.3°, resulting in strongly AFM contributions. As for $J_2$ interaction, although two cation-anion-anion angles of 132.1° and 89.5° contribute to AFM and FM respectively, the resulting extended $J_2$ still exhibits AFM sign due to the AFM dominance in monolayer MnPSe$_3$ [44]. Besides, compared with $J_1$ and $J_3$, $J_2$ is relatively weak for involving small Se-Se hybridizations. Hence, we suppose that the AFM-FM transition may mainly depend on the significant competition effect between AFM exchange interaction and FM superexchange interaction from the NN interaction $J_1$. With the increase of biaxial strain, the cation-anion-anion angels always change within the criterion angle range. Values of $J_2$ and $J_3$ are expected to reduce in magnitude because of the increasing atomic distances, while the variation of $J_1$ is subtler. A rapid reduction of the direct exchange interaction and a relatively slow reduction of the indirect superexchange interaction will give rise to the impairment of AFM stability and resultant enhancement of FM stability. Once the tensile strain is large enough, the magnetic phase reversal can be realized by changing the signs (positive and negative) of the energy difference between AFM and FM, and thus the AFM-FM transition can be well explained.

Additionally, the uniaxial strain effect on magnetic phase transition has been recognized in previous CrPS$_4$ monolayer [45]. In this work, uniaxial strain engineering is further tested. The results are listed in Table S2 in the Supplementary Information. Generally, both the uniform biaxial and uniaxial strains could trigger the modification in total energies and result in the magnetic phase transition. However, the effect is weaker for uniaxial strain in monolayer MnPSe$_3$. A relatively large uniaxial strain is required to alter the AFM-FM transition.

## 4. Conclusion

In summary, the electronic structures and magnetic properties of the strained monolayer MnPSe$_3$ are systematically surveyed. Our *ab initio* calculations show that the 2D MnPSe$_3$ is an AFM semiconductor. In particular, the AFM stability can be substantially influenced by strain engineering and an AFM to FM transition occurs under a large tensile strain, indicating that the strain can be used as an effective knob for modulating the magnetic phase. The underlying physical mechanism of strain-dependent magnetic stability is further elucidated as the competition effect of direct exchange and indirect superexchange interactions between the NN two Mn atoms. The research discussed here provides strong evidence that strain can be an effective pathway to induce or modulate magnetic properties in 2D crystal and sheds light on the potential candidate of MnPSe$_3$ as a promising 2D magnet.

# Acknowledgements

This work is supported by National Natural Science Foundation of China (51671142, U1632152 and 51661145026) and Key Project of Natural Science Foundation of Tianjin City (16JCZDJC37300).

**Figure captions**

**Fig. 1.** (a) Top and side views of monolayer MnPSe$_3$. The unit cell is indicated by red solid line. The grey, purple and green spheres represent the P, Mn, and Se atoms, respectively. (b) Atomic structures of a distorted MnSe$_6$ octahedron and [P$_2$Se$_6$]$^{4-}$ bipyramids enclosed with hexagonal Mn atoms. Isosurface plots of the spin charge density for (c) FM and (d) AFM configurations. The isosurface value is set up to ±0.1 e/Å$^3$. The blue and red isosurfaces represent spin-up and spin-down charge densities, respectively. (e) Four possible spin configurations and their relative energy Δ*E* (meV/unitcell) with respect to the Néel-AFM configuration.

**Fig. 2.** (a) Variation of the energy difference between the AFM and FM coupling and the nearest Mn-Mn distance in biaxial strained systems. Negative value indicates that the FM configuration is less stable than the AFM configuration. (b) Variation of the band gap at different biaxial strains. (c) Schematic plot of the effect of biaxial strain on the magnetic coupling.

**Fig. 3.** Band structures and density of states for monolayer MnPSe$_3$ at the biaxial strain of (a) -5%, (b) 0%, (c) 5%, (d) 10%, (e) 13% and (f) 15%. Fermi level is indicated by the vertical solid line and set to zero.

**Fig 4.** Schematic diagram of possible paths for the (a) NN, (b) second NN and (c) third NN magnetic exchange interactions in monolayer MnPSe$_3$.

**Fig. 1, Q. Pei *et al.***

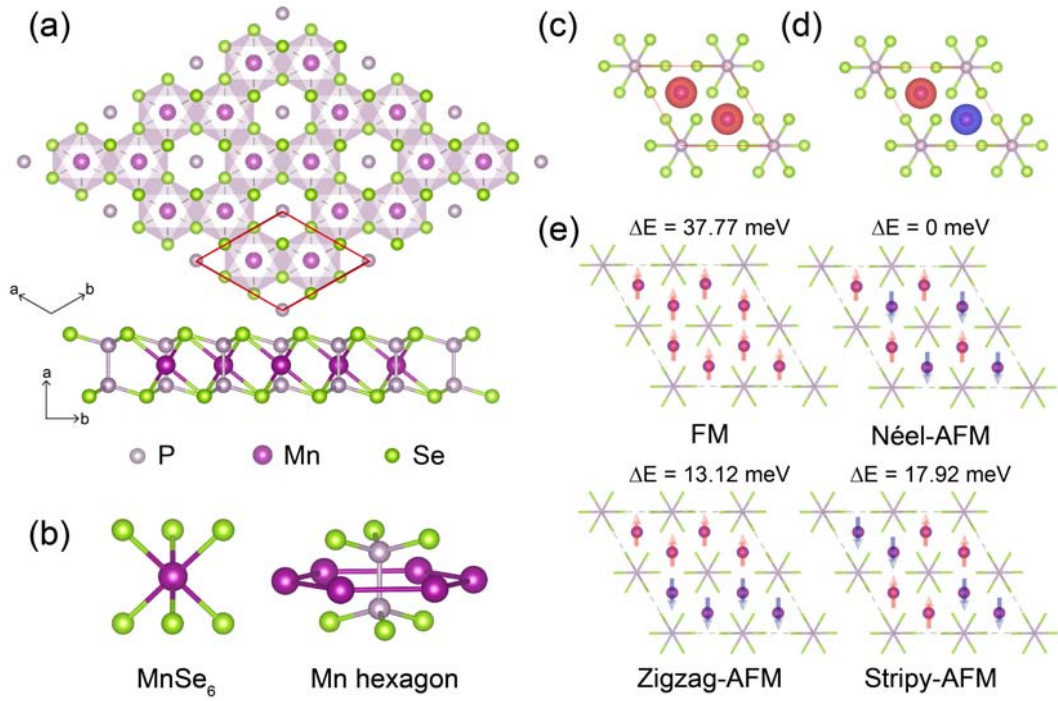

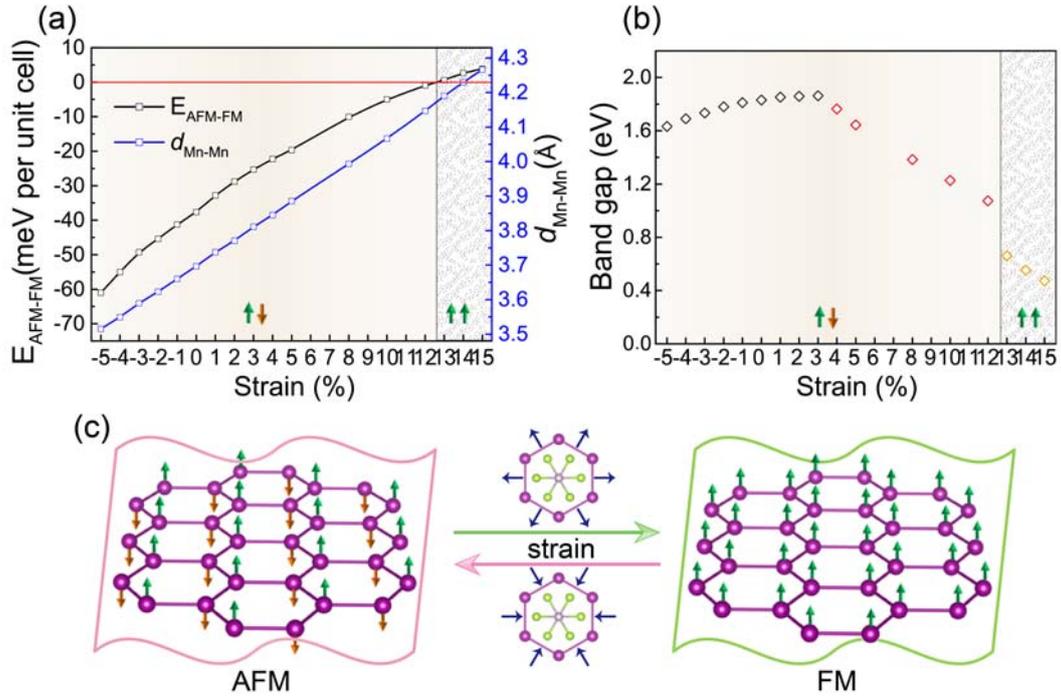

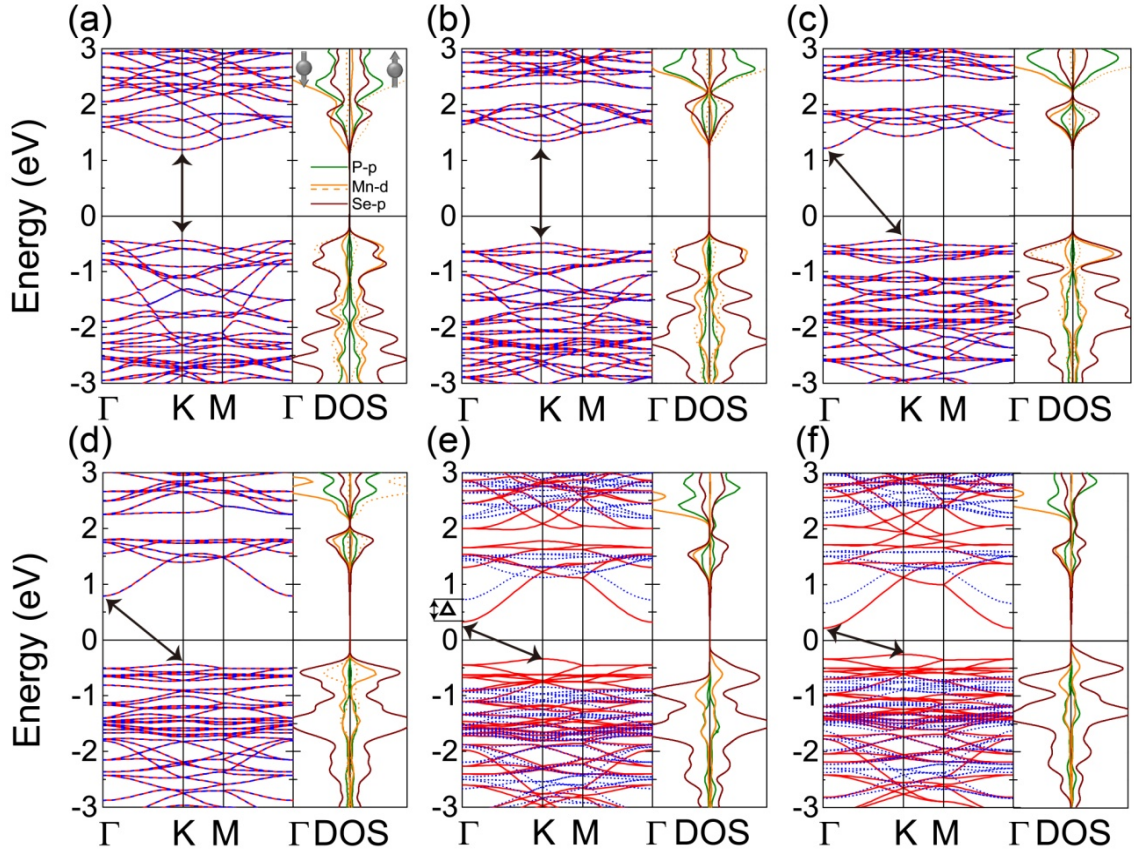

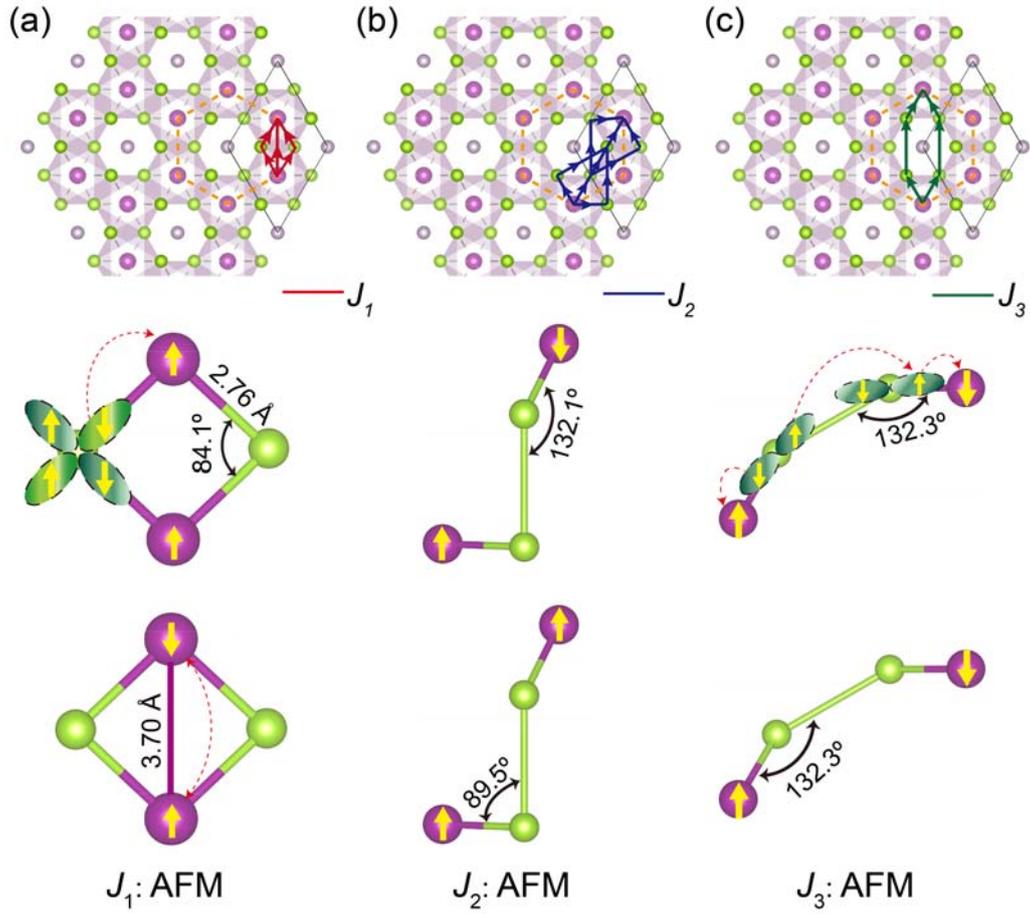

# Supplementary Information

# Tunable Electronic Structure and Magnetic Coupling in Strained Two-Dimensional Semiconductor MnPSe$_3$

Qi Pei[1], Xiaocha Wang[2], Jijun Zou[3], Wenbo Mi[1,*]

[1]*Tianjin Key Laboratory of Low Dimensional Materials Physics and Preparation Technology, School of Science, Tianjin University, Tianjin 300354, China*

[2]*School of Electrical and Electronic Engineering, Tianjin University of Technology, Tianjin 300384, China*

[3]*Key Laboratory for Green Chemical Technology of the Ministry of Education, School of Chemical Engineering and Technology, Tianjin University, Tianjin 300354, China*

[*]Author to whom all correspondence should be addressed.

E-mail: miwenbo@tju.edu.cn

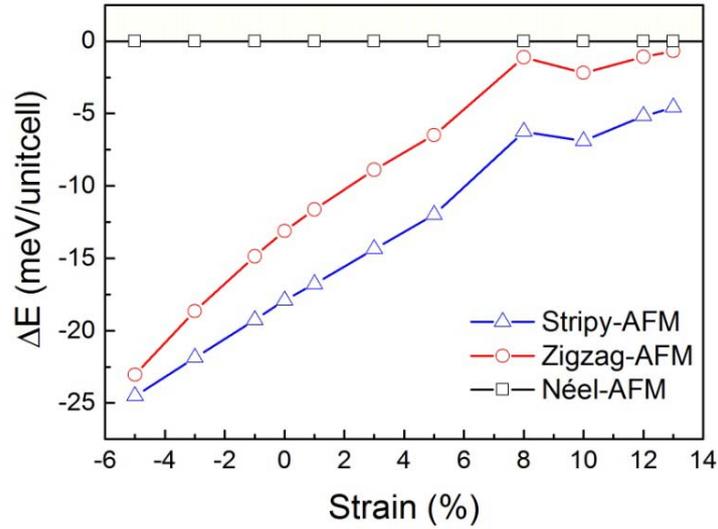

**Fig. S1** Variation of total energy relative to the Néel-AFM configuration among Néel-AFM, zigzag-AFM, stripy-AFM. Negative value indicates that zigzag-AFM and stripy-AFM configurations are less stable than the Néel-AFM configuration.

Fig. S1 shows the variation of the energy difference between the regular AFM (Néel-AFM) and other AFM configurations (stripy-AFM and zigzag-AFM) for biaxial strained monolayer MnPSe$_3$. Negative values indicate that other AFM configurations are less stable than the regular AFM configuration. Within the biaxial strain range from -5% to 13%, the values of Δ$E$ almost increases monotonically with biaxial strain. When the biaxial strain increases to 13%, Δ$E_{\text{Néel-zigzag}}$ is relatively smaller with the value of -0.69 meV. However, the spins of Mn atoms with Néel order always possess the lowest energy in the three AFM arrangements, showing a structural stability. Hence, within the biaxial strain range we considered, there is neither a magnetic phase transition from Néel to other AFM phases nor a transition from other AFM phases to FM phase.

**Table S1** Lattice constant $a$ (Å), magnetic ground state (GS) and exchange coupling constants (meV) in pristine and biaxial strained monolayer MnPSe$_3$. The *ab* initio calculations are performed within GGA+$U$+D2 ($U$=5 eV for Mn-3$d$ orbital).

|  | $a$ (Å) | GS | $J_1$ (meV) | $J_2$ (meV) | $J_3$ (meV) |
|---|---|---|---|---|---|
| MnPSe$_3$ | 6.403 | Néel | 0.195 | 0.020 | 0.103 |
| MnPSe$_3$ (1% strain) | 6.467 | Néel | 0.159 | 0.014 | 0.095 |
| MnPSe$_3$[16] | 6.334 | Néel | 0.231 | 0.021 | 0.141 |

**Table S2** Lattice constant (Å), magnetic ground state (GS) and total energy (meV/unitcell) relative to the Néel-AFM configuration among Néel-AFM, zigzag-AFM, stripy-AFM and FM configurations in uniaxial strained monolayer MnPSe$_3$.

| Strain (%) | $a$ (Å) | $b$ (Å) | GS | Néel (meV) | Zigzag (meV) | Stripy (meV) | FM (meV) |
|---|---|---|---|---|---|---|---|
| -5% | 6.083 | 6.403 | Néel | 0 | -23.84 | -20.82 | -49.99 |
| -3% | 6.211 | 6.403 | Néel | 0 | -18.98 | -19.86 | -44.30 |
| -1% | 6.339 | 6.403 | Néel | 0 | -14.81 | -18.60 | -39.77 |
| 1% | 6.467 | 6.403 | Néel | 0 | -11.64 | -17.29 | -36.12 |
| 3% | 6.595 | 6.403 | Néel | 0 | -9.503 | -15.96 | -33.37 |
| 5% | 6.723 | 6.403 | Néel | 0 | -7.988 | -14.43 | -31.09 |
| 7% | 6.851 | 6.403 | Néel | 0 | -7.025 | -12.54 | -28.82 |
| 10% | 7.04 | 6.403 | Zigzag | 0 | **0.433** | -7.958 | -24.71 |

Table S2 lists the lattice constant, GS and total energy relative to the Néel-AFM configuration among Néel-AFM, zigzag-AFM, stripy-AFM and FM configurations in uniaxial strained monolayer MnPSe$_3$. Generally, both the uniform biaxial and uniaxial strains could trigger the modification in total energies of different magnetic phases and result in the magnetic phase transition. However, in monolayer MnPSe$_3$, the effect is weaker for uniaxial strain by comparing with the variation of total energies in biaxial strained system. With the uniaxial strain changing from -5% to 10%, the energy difference between AFM and FM configurations only changes from 49.99 to 24.71 meV. Hence, a relatively large uniaxial strain is required to alter the AFM-FM transition. Additionally, we notice a clear transition from Néel-AFM to zigzag-AFM when the uniaxial strain increases to 10%, which can be ascribed to the small energy difference between the two magnetic phases.